# OBSERVATION OF INSTABILITIES OF COHERENT TRANSVERSE OSCILLATIONS IN THE FERMILAB BOOSTER*

Y. Alexahin#, N. Eddy, E. Gianfelice-Wendt, V. Lebedev, W. Pellico, W. Marsh, K. Triplett, FNAL, Batavia, IL 60510, U.S.A.

*Abstract*

The Fermilab Booster - built more than 40 years ago - operates well above the design proton beam intensity of $4 \cdot 10^{12}$ ppp. Still, the Fermilab neutrino experiments call for even higher intensity exceeding $5.5 \cdot 10^{12}$ ppp. A multitude of intensity related effects must be overcome in order to meet this goal including suppression of coherent dipole instabilities of transverse oscillations which manifest themselves as a sudden drop in the beam current. In this report we present the results of observation of these instabilities at different tune, coupling and chromaticity settings and discuss possible cures.

## INTRODUCTION

For more than four decades the Fermilab Booster [1] has been the cornerstone of the Fermilab accelerator complex and will retain this role for the foreseeable future. The basic parameters of the present Booster operation are given in Table 1.

Table 1: Booster basic parameters

| Parameter | Unit | Value |
|---|---|---|
| Kinetic energy (injection/final) | GeV | 0.4/8 |
| Circumference | m | 474.25 |
| Transition $\gamma_t$ | - | 5.48 |
| RF harmonic number | - | 84 |
| Protons/batch | | $4.5 \cdot 10^{12}$ |
| Magnet cycle frequency | Hz | 15 |
| Average repetition rate | Hz | 9 |

The Booster significantly surpassed the design goals, but the next generation of Fermilab neutrino experiments require even higher proton output, $\sim 6 \cdot 10^{12}$ ppp at 15Hz. To make such an increase possible it is necessary to understand and suppress coherent instabilities as well as to minimize beam losses especially in the later part of the ramp.

According to recent theoretical results [2] the largest transverse impedances in the Booster are introduced by laminated magnets and can drive fast instabilities with growth rates approaching 0.01 turn$^{-1}$ at injection energy.

In this paper we report some observations of coherent instabilities made mostly in the course of development of a console application for turn-by-turn measurements of tunes, coupling and chromaticity [3]. Full-scale dedicated studies of these phenomena at the Booster will be performed in the future.

## STABILIZING MECHANISMS

The main "knob" for transverse instability control in bunched beams is chromaticity which non-trivially modifies Landau damping and brings about the head-tail damping.

### Landau Damping

Landau damping takes place when the coherent tune falls within the range of incoherent tunes. The spread in incoherent tunes can be produced by lattice nonlinearities as well as by space charge forces.

However, the space charge also produces a shift of incoherent tunes which is always larger than its contribution to the tunespread so the space charge is usually considered as a destabilizing factor. But in a bunched beam the situation can be different due to strong modulation of the space charge tuneshift in the course of synchrotron oscillations [4].

Apart from this possible intervention of space charge forces, Landau damping has weak – if any – dependence on the beam intensity. Since the instability growth rate is proportional to intensity there is a threshold intensity above which Landau damping is insufficient to stabilize the beam.

### Head-Tail Damping

Interaction of the tail particles with the wake field left by the head of the bunch and their interchange lead to the so called head-tail effect (see e.g. [5]). It redistributes damping/growth rates of different head-tail modes leaving the total sum equal to zero. If the chromaticity and the slippage factor $\eta$ are of the same sign (e.g. if chromaticity is negative below transition energy) it provides damping of the lowest mode with rate

$$\alpha_y^{(HT)} \sim \frac{I_b Q_y' |W_y|}{E\eta} > 0, \qquad (1)$$

where $I_b$ is the bunch current, $W_y$ is dipole wake function (assumed negative at distances of interest). The total energy $E$ is assumed to be far enough from $mc^2\gamma_t$ to avoid complications with small $|\eta|$.

Since the damping rate (1) is proportional to the beam intensity there is no threshold for the lowest mode stability, however, higher modes require Landau damping for stability which may be lost at high intensities [5].

___________________________________________
* Work supported by Fermi Research Alliance, LLC under Contract DE-AC02-07CH11359 with the U.S. DOE.
#alexahin@fnal.gov

## Effects of Coupling

According to R. Talman's conjecture [6] in the presence of transverse coupling the stability condition (1) should include chromaticities and wakes in both planes so that it may not be necessary to fulfil it for each plane separately. Basing on theory of coupled motion a more rigorous estimate of normal mode damping rate was obtained in [7]:

$$\alpha_n^{(HT)} \sim \frac{I_b Q_n'}{E\eta} \langle \beta_{xn}|W_x| + \beta_{yn}|W_y| \rangle > 0 \qquad (2)$$

where $\beta_{xn}$ and $\beta_{yn}$ are beta-functions of coupled motion describing projection of mode $n$ on respective planes. Coupling strongly affects the normal mode chromaticities $Q_n'$ resulting in the so-called *chromaticity sharing* effect which was confirmed experimentally [6].

## Transverse Dampers

All the above methods of coherent motion stabilization may lead to higher losses and emittance growth due to incoherent effects, so the feedback systems were added for each plane.

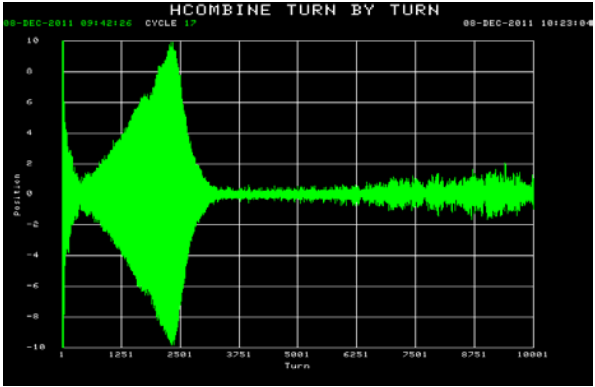

Figure 1: Combined TBT signal from HBPMs (arbitrary units) at $N_p = 4 \cdot 10^{12}$ after coupling correction.

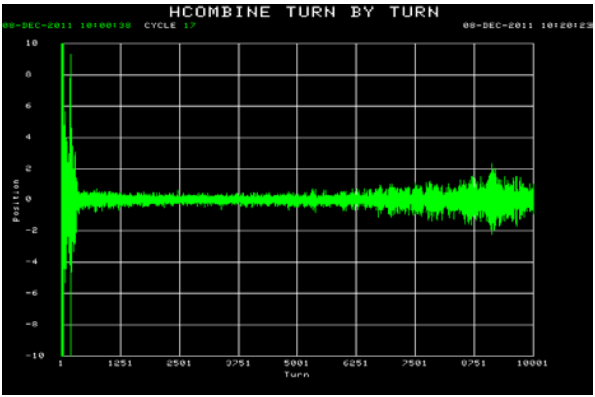

Figure 2: TBT signal under the same conditions as for Fig. 1 with the horizontal damper on.

The damper system consists of a four plate stripline pickup to determine the horizontal and vertical displacement of each bunch, a digital damper board to process the bunch by bunch signals, and two stripline kickers, one in each plane, to kick each bunch [8]. The digital damper is able to record the positions for each bunch throughout the Booster ramp. This feature allows analyzing the bunch by bunch behaviour during the instabilities.

Effectiveness of the dampers (horizontal in the particular case) is illustrated by Figs. 1 and 2 which show the turn-by-turn (TBT) BPM signal under unstable conditions with damper off and on.

## INSTABILITY OBSERVATIONS

Until recently the Booster operated with rather strong coupling, $C \approx 0.06$, in the early part of the ramp about 3000 turns or 6 ms from injection [3]. Attempts to correct this strong coupling lead to coherent instability.

One would expect the vertical motion to be more prone to instability since the vertical wake fields in the Booster are stronger:
- inside the magnets (where there is no beam pipe) the horizontal aperture is significantly larger so that $W_y \approx 2W_x$,
- the strongest kicker is the vertical extraction kicker which is stronger coupled to the vertical motion,
- in RF cavities $\beta_y \approx 3\beta_x$ so again the HOM are expected to affect the vertical motion more.

Still, after coupling was reduced to $C < 0.01$ and the tunes set apart ($Q_x \approx 6.78$, $Q_y \approx 6.85$) it turned out that the instability was horizontal with a rather low growth rate $\sim 5 \cdot 10^2$ s$^{-1}$ (Fig. 1).

### Horizontal Instability

To study the bunch-by-bunch instability pattern the horizontal damper pickup was used. Figure 3 shows that multi-bunch mode 8 was excited. The first two bunches are missing to create the extraction gap, the first period after the gap having smaller amplitude of oscillations.

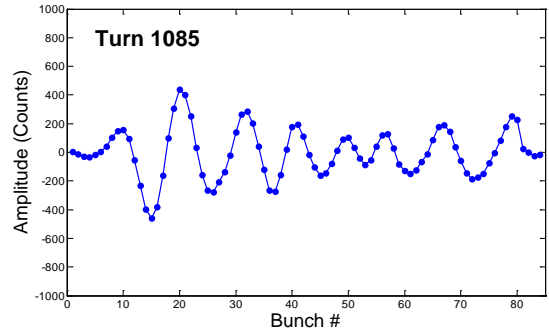

Figure 3: Bunch-by-bunch horizontal positions at the onset of horizontal instability

The revolution frequency at this energy is $f_0$=0.47MHz, so the lowest frequency of the signal $f_{min}$ is close to 3.8MHz. The instability may be caused by coupling to a resonant element with eigenfrequency $Mhf_0 \pm f_{min}$, where $h$=84 is the RF harmonic number and $M$=0,1,2,… It is not clear yet what this element might be.

Subsequent studies showed that without coupling the horizontal chromaticity should be increased by absolute value (Fig. 4), whereas the vertical chromaticity can be reduced to ~1/6 of the horizontal value in a striking disagreement with our expectations.

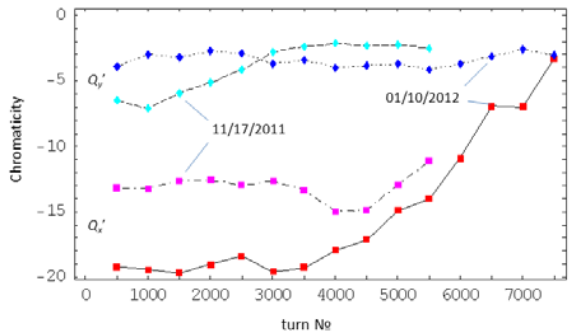

Figure 4 (color): Chromaticities in the early part of the ramp as found after coupling correction (11/17/2011) and as required for beam stability (01/10/2012).

Since the horizontal chromaticity was already higher than the vertical one before coupling correction, it was not the chromaticity sharing that provided beam stability. Eq. (2) shows that what matters is the product of chromaticity and wake functions and therefore suggests another possibility which may be called *wake-field sharing*: in the presence of coupling the larger $W_y$ helps stabilizing the horizontal motion too, while when coupling is corrected the lack of this contribution must be compensated by a larger negative horizontal chromaticity.

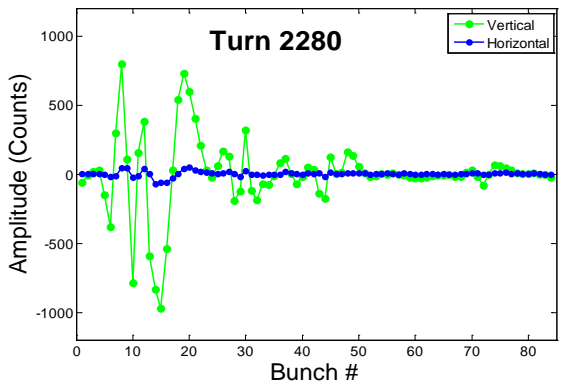

Figure 5 (color): Bunch-by-bunch horizontal and vertical positions at the onset of vertical instability.

### Vertical Instability

At chromaticities as low as $Q_y' \approx -3$ the vertical motion also becomes unstable but in a quite different way (Fig. 5). Approximately 20 bunches in the head of the beam are strongly affected by the instability which has no apparent bunch-by-bunch pattern. The characteristic frequency is significantly higher than in the horizontal case, so the vertical instability most likely has a different source.

### Present Status

After finding the conditions for stable operation below transition, chromaticity above transition energy was also corrected [2] permitting to noticeably increase beam intensity. Figure 6 shows the number of protons in the batch and the integrated dose at two beam loss monitors over the full Booster ramp for three values of initial intensity (13, 14 and 15 turns injection).

The goal of $5.5 \cdot 10^{12}$ ppp was achieved with 14 turns injection but at the price of high losses, mostly after injection but also after transition crossing.

The crossing happens at about 9500 turns after injection, close to the end of range of Figs. 1 and 2. One can see some activity going on there which was also detected in the vertical plane. The BPMs provide integrated signal over a number of bunches so the maximum amplitude can be much higher. We are planning to address this issue as well as to look for the source of the horizontal instability in the future.

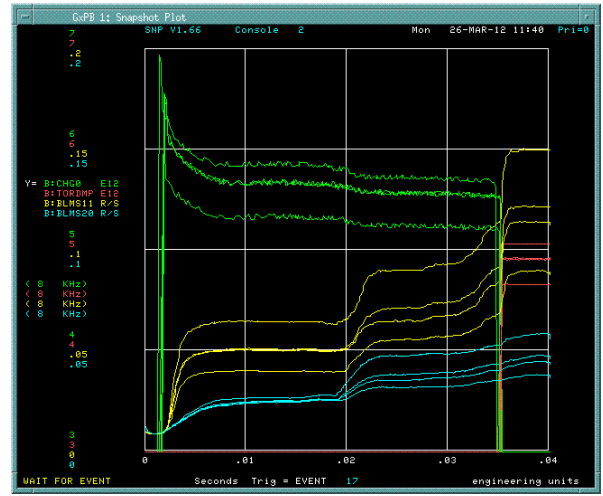

Figure 6 (color): Total number of protons and the integrated dose at two BLMs vs. time for three values of initial intensity.